%% file: main.tex
\documentclass[conference]{IEEEtran}

\usepackage{enumitem}

\usepackage[pdftex]{graphicx}
\graphicspath{{../pdf/}{../jpeg/}}
\DeclareGraphicsExtensions{.pdf,.jpeg,.png}
\usepackage{tikz}
\usepackage{xcolor} 
\usepackage[cmex10]{amsmath}
\usepackage{siunitx}
\usepackage{soul}
\usepackage{eqparbox}
\usepackage{url}
\usepackage{booktabs}
\usepackage{multirow}
\usepackage[nocompress]{cite}
\usepackage{tabularx}
\newcolumntype{Y}{>{\raggedright\arraybackslash}X}

\usepackage{amssymb}
\usepackage{textgreek}
\usepackage{algorithm}
\usepackage{algpseudocode}  

\usepackage{lettrine}

\begin{document}
\addtolength{\topmargin}{0.05in}

\title{
\vspace{-1cm}
\begin{tikzpicture}[remember picture, overlay]
\node[anchor=north, yshift=-0.6cm] at (current page.north) {%
\fbox{\parbox{0.97\textwidth}{\centering\footnotesize
This paper has been accepted for publication at the IEEE International Conference on Communications (ICC 2026). Copyright may be transferred without notice, after which this version may no longer be accessible. This is the camera-ready version.
}}};
\end{tikzpicture}
\vspace{0.4cm}

AI-Driven Fuzzing for Vulnerability Assessment of 5G Traffic Steering Algorithms
}

\author{
    \IEEEauthorblockN{
        Seyed Bagher Hashemi Natanzi\IEEEauthorrefmark{1},
        Hossein Mohammadi\IEEEauthorrefmark{2},
        Bo Tang\IEEEauthorrefmark{1},
        Vuk Marojevic\IEEEauthorrefmark{2}
    }
    \IEEEauthorblockA{
        \IEEEauthorrefmark{1}Electrical and Computer Engineering Department, Worcester Polytechnic Institute, USA\\
        \IEEEauthorrefmark{2}Electrical and Computer Engineering Department, Mississippi State University, USA\\
        Email: \texttt{\{snatanzi, btang1\}@wpi.edu}, \texttt{\{hm1125, vm602\}@msstate.edu}
    }
}
\definecolor{lime}{HTML}{A6CE39}
\DeclareRobustCommand{\orcidicon}{%
    \begin{tikzpicture}
    \draw[lime, fill=lime] (0,0) 
    circle [radius=0.16] 
    node[white] {\tiny ORCID}; 
    \draw[white, fill=white] (-0.0625,0.095) 
    circle [radius=0.007];
    \end{tikzpicture}
    \hspace{-2mm}
}

\newcommand{\orcidauthor}[2]{%
    #1\,\orcidicon 
}

\maketitle

\input{include/abstract}
\input{include/Introduction}

\input{include/Background}
\input{include/System_Model_and_Methodology}

\input{include/Proposed_Method}

\input{include/Simulation_Results_and_Analysis}

\input{include/Conclusion}

\section*{\textcolor{black}{Acknowledgment}}
\noindent
\footnotesize This work was supported in part by NTIA Awards 51-60-IF007 and 28-60-IF012 and NSF Award CNS-2120442. The views expressed are those of the authors and do not necessarily represent those of the NTIA or NSF.

\enlargethispage{\baselineskip}
\bibliographystyle{IEEEtran}
\bibliography{./bib/main.bib}
\end{document}

%% file: include/abstract.tex
\begin{abstract}
Traffic Steering (TS) dynamically allocates user traffic across cells to enhance Quality of Experience (QoE), load balance, and spectrum efficiency in 5G networks. However, TS algorithms remain vulnerable to adversarial conditions such as interference spikes, handover storms, and localized outages. To address this, an AI-driven fuzz testing framework based on the Non-Dominated Sorting Genetic Algorithm II (NSGA-II) is proposed to systematically expose hidden vulnerabilities. Using NVIDIA Sionna, five TS algorithms are evaluated across six scenarios. Results show that AI-driven fuzzing detects 34.2\% more total vulnerabilities and 5.8\% more critical failures than traditional testing, achieving superior diversity and edge-case discovery. The observed variance in critical failure detection underscores the stochastic nature of rare vulnerabilities. These findings demonstrate that AI-driven fuzzing offers an effective and scalable validation approach for improving TS algorithm robustness and ensuring resilient 6G-ready networks.
\end{abstract}

\begin{IEEEkeywords}
Traffic steering, AI-driven fuzzing, vulnerability assessment, NSGA-II, network resilience.
\end{IEEEkeywords}

%% file: include/Introduction.tex
\section{Introduction}

The dynamic nature of 5G networks demands advanced algorithms to manage user equipment (UE) assignments across base stations (BSs), ensuring efficient spectrum utilization and consistent Quality of Experience (QoE)~\cite{10742579}. Central to this functionality is traffic steering (TS), which dynamically optimizes UE–gNodeB associations to maintain load balance and network efficiency. Despite their sophistication, TS algorithms remain susceptible to adversarial conditions, severe interference, localized outages, and operational failures such as Evolved Packet System (EPS) fallback failure. These failures often stem from rapid radio fluctuations, inaccurate network-state estimation, or delayed control signaling under heavy load~\cite{10102369, 10.1109/TWC.2024.3396273}, and are difficult to prevent due to the stochastic nature of wireless environments. Handover failures and suboptimal load-balancing decisions are frequent triggers with cascading impacts on QoE, spectrum efficiency, and operator profitability potentially leading to measurable revenue losses in dense urban deployments.

Traditional Testing approaches, typically limited to static or traditional scenarios, fail to reveal the full breadth and depth of such flaws. This limitation underscores the need for advanced automated methodologies to rigorously assess algorithmic resilience in realistic and adversarial network conditions~\cite{mayeke2025evaluating}.

To address this gap, this study proposes an artificial intelligence (AI)-driven fuzzing framework based on NSGA-II to generate adversarial network states that expose rare TS failures affecting QoE, fairness, and stability. The approach outperforms traditional testing by 34.2\% in overall vulnerability detection, with \textit{p}~$<$~0.00001 indicating a highly significant improvement unlikely due to chance. It further provides architecture-level insights, showing that rule-based TS policies are more prone to failures under adversarial conditions than adaptive learning-based agents. These results establish AI-driven fuzzing as a statistically validated and architecture-aware method for enhancing the robustness of 5G TS systems.

The remainder of this paper is structured as follows: Section~\ref{sec:Background} surveys related work, Section~\ref{sec:system_model_methodology} elaborates on the system model and methodology, Section~\ref{sec:results} presents the evaluation results, and Section~\ref{sec:conclusion} provides conclusions and future research directions.

%% file: include/Background.tex
\section{Background and Related Work}
\label{sec:Background}

Traffic Steering in 5G networks encompasses three primary categories: mobility-based steering for optimizing handovers, load-based steering for balancing network resources, and QoS-based steering for differentiating services. Classic approaches include 3GPP-standardized mechanisms such as Event A3-based handover and Cell Individual Offset (CIO) adjustments, which rely on predefined Reference Signal Received Power (RSRP) thresholds. Modern implementations increasingly adopt machine learning (ML) techniques, from supervised learning for pattern prediction to reinforcement learning for adaptive decision-making~\cite{10102369, MUNOZ2014100}.
Despite extensive research on TS optimization~\cite{10102369}, vulnerability assessment remains largely unexplored. Existing studies focus on maximizing throughput or minimizing handover rates under normal conditions, overlooking failure modes that emerge under stress. For instance, reinforcement learning shows promise for adaptive control, but often fails catastrophically when encountering out-of-distribution scenarios not seen during training.

A \textit{vulnerability} denotes network states where performance degrades beyond acceptable limits (per 3GPP TS 22.261), affecting stability, QoE, or fairness.
Current testing methodologies for TS algorithms fall into three categories: \textit{Conventional testing}, which uses predefined scenarios but misses edge cases, \textit{formal verification}, which provides theoretical guarantees but becomes computationally intractable for realistic 5G models with hundreds of state variables~\cite{10729871}, and \textit{fuzzing techniques}, which have shown success for protocol testing~\cite{she2019neuzz, yang2024systematic, sun20245gc} and which focus on implementation errors rather than on algorithmic vulnerabilities specific to TS logic.
The gap between optimization-focused TS research and security-oriented fuzzing presents an opportunity. Multi-objective genetic algorithms have explored network tradeoffs~\cite{goudos2018multi}, but have not been applied for vulnerability discovery. 
This study bridges this gap by adapting NSGA-II for TS vulnerability assessment, introducing domain-specific fitness functions that simultaneously capture QoE degradation, fairness violations, and stability issues. Beyond academic relevance, the proposed framework also provides practical utility for telecom vendors’ back-office teams by enabling systematic identification of TS vulnerabilities during testing and validation.

%% file: include/System_Model_and_Methodology.tex
\section{System Model and Problem Formulation}
\label{sec:system_model_methodology}

This section defines the analytical system model for the 5G Traffic Steering (TS) environment and formulates the vulnerability discovery problem as a multi-objective optimization task. The formulation establishes the basis for the proposed NSGA-II–based AI-driven fuzzing framework.

\subsection{System Model}

We consider a 5G network composed of a set of $M$ gNodeBs $\mathcal{B}=\{b_1,b_2,\ldots,b_M\}$ and a set of $N$ UEs $\mathcal{U}=\{u_1,u_2,\ldots,u_N\}$. Each UE $u_i\!\in\!\mathcal{U}$ is associated with exactly one gNodeB $b_j\!\in\!\mathcal{B}$ at any given time~$t$.

The overall network state at time $t$ is denoted as
\[
S_t = \{P_t,L_t,A_t,C_t\},
\]
where
\begin{itemize}
    \item $P_t\in\mathbb{R}^{N\times 2}$ represents the UE positions in the 2D plane,
    \item $L_t\in[0,1]^M$ is the normalized cell-load vector, where $L_j = n_j/N_j^{\max}$ with $n_j$ being the number of active UEs in cell $j$,
    \item $A_t\in\{0,1\}^{N\times M}$ denotes the association matrix, where $a_{ij}=1$ if $u_i$ is connected to $b_j$ and $0$ otherwise, and
    \item $C_t\in\mathbb{R}^{N\times M}$ represents the channel quality matrix (RSRP or SINR) between every UE–gNodeB pair.
\end{itemize}

For each UE-to-gNodeB link, the received signal-to-interference-plus-noise ratio (SINR) is modeled as
\begin{equation}
\text{SINR}_{i,j} = \frac{P_j G_{i,j}}{I_i + N_0B},
\end{equation}
where $P_j$ is the transmit power of $b_j$, $G_{i,j}$ is the channel gain (including path loss, shadowing, and fading), $I_i$ is the aggregate inter-cell interference, $B$ is the system bandwidth, and $N_0$ is the noise power spectral density.

The throughput of UE $u_i$ connected to $b_j$ is approximated as
\begin{equation}
T_{i,j} = B \log_2(1+\text{SINR}_{i,j}),
\label{eq:throughput_def}
\end{equation}
and the cell load $L_j$ evolves according to the active-UE occupancy ratio.

A TS algorithm $\pi: S_t \rightarrow A_t$ maps network states to association decisions, determining UE assignments to gNodeBs based on criteria such as received power, cell load, and QoS demand.

\subsection{Problem Formulation}

The vulnerability discovery problem aims to identify network configurations that trigger operational failures beyond acceptable limits. These limits are abstractly defined as
\[
\begin{aligned}
\text{Stability failure:} &\quad H_{\text{rate}} > \tau_{\text{stab}},\\
\text{QoE degradation:} &\quad \text{Thr}_{5\%} < \tau_{\text{QoE}},\\
\text{Fairness violation:} &\quad J < \tau_{\text{fair}},
\end{aligned}
\]
where $H_{\text{rate}}$ denotes the average handover rate per UE, $\text{Thr}_{5\%}$ is the 5th-percentile user throughput, and $J$ is Jain’s fairness index. The parameters $\tau_{\text{stab}}$, $\tau_{\text{QoE}}$, and $\tau_{\text{fair}}$ represent operator-defined thresholds aligned with 3GPP TS 22.261 service requirements to ensure operational relevance.

Given a TS policy $\pi$, the goal is to identify network states that maximize the exposure of vulnerabilities through multi-objective optimization:
\begin{equation}
\max_{z\in\mathcal{Z}} F(z) = [f_1(z),f_2(z),f_3(z)], \label{eq:obj_vector}
\end{equation}
where $z$ encodes the candidate network configuration, and the objectives target instability, QoE degradation, and unfairness:
\begin{align}
f_1(z) &= \frac{1}{N} \sum_{i=1}^{N} \mathcal{I}(c_i(z) \neq c_i^{\text{prev}}) \quad &\text{(Instability)}, \label{eq:f1}\\[3pt]
f_2(z) &= \frac{1}{\text{Thr}_{5\%}(z) + \epsilon} \quad &\text{(QoE degradation)}, \label{eq:f2}\\[3pt]
f_3(z) &= 1 - \frac{\left(\sum_{i=1}^{N} T_i(z)\right)^2}{N \cdot \sum_{i=1}^{N} T_i(z)^2} \quad &\text{(Unfairness)}. \label{eq:f3}
\end{align}

The formulation in~(\ref{eq:obj_vector})--(\ref{eq:f3}) ensures simultaneous coverage of 5G TS failure modes. Objective $f_1$ captures ping-pong instability, while $f_2$ targets QoE degradation using the base-2 logarithm throughput defined in (\ref{eq:throughput_def}). Objective $f_3$ measures unfairness via Jain's Fairness Index. These objectives are evaluated against operator-defined thresholds $\tau_{\text{stab}}$, $\tau_{\text{QoE}}$, and $\tau_{\text{fair}}$, aligned with 3GPP TS 22.261 requirements, to robustly identify operational vulnerabilities.

%% file: include/Proposed_Method.tex
\section{Proposed Method}
\label{sec:proposed_method}
The proposed design enables Pareto-front exploration through the well-established NSGA-II algorithm without ad-hoc scalarization. The methodology employs NSGA-II as the core optimization engine and integrates it with a realistic 5G simulation environment based on the Sionna library~\cite{sionna}. The complete implementation is available online\footnote{https://github.com/CLIS-WPI/AI-Fuzzing}. The goal is to identify adversarial network configurations that expose failures in TS algorithms while maintaining computational efficiency and solution diversity across the Pareto front. The algorithm uses selection, crossover, and mutation operators to evolve candidate configurations while preserving diversity via crowding distance and elitism.

\subsection{NSGA-II-Based Vulnerability Discovery Engine}

We formulate the vulnerability discovery process as a three-objective search problem over the space of network configurations. NSGA-II is employed to approximate the Pareto-optimal set without ad-hoc scalarization. Let $\mu$ denote the population size and $G$ the number of generations. At generation $g$, the population $\mathcal{P}_g$ consists of $\mu$ candidate configurations $z_i$, each encoding parameters such as UE positions, cell loads, and link conditions. The algorithm iteratively evolves $\mathcal{P}_g$ through the following operators:

\begin{itemize}
    \item \textbf{Selection:} Tournament selection based on dominance rank, $\text{rank}(z_i)$, and crowding distance $d(z_i)$ to preserve diversity.
    \item \textbf{Crossover:} A blend crossover operator with probability $p_c$ combines parent vectors $z_i$ and $z_j$ to generate offspring $z_k$.
    \item \textbf{Mutation:} Gaussian or uniform mutation with probability $p_m$ perturbs selected variables, introducing exploration within the configuration space. The mutation intensity is governed by variance $\sigma$.
\end{itemize}

Each candidate configuration $z_i$ is evaluated through the objective vector $F(z_i)=[f_1,f_2,f_3]$ defined in~(\ref{eq:obj_vector})–(\ref{eq:f3}), capturing handover instability, QoE degradation, and unfairness. Non-dominated sorting and elitism ensure that superior solutions persist across generations, while the crowding distance metric maintains front diversity.

Algorithm~\ref{alg:nsga2} summarizes the iterative optimization procedure.

\begin{algorithm}[t]
\caption{NSGA-II for Vulnerability Discovery}
\label{alg:nsga2}
\begin{algorithmic}[1]
\State Initialize population $P_0$ with $\mu$ random configurations
\For{$g = 1 \to G$}
    \State Apply tournament selection on $P_g$
    \State Generate offspring $Q_g$ via blend crossover (rate $p_c$)
    \State Apply Gaussian mutation to $Q_g$ (rate $p_m$, variance $\sigma$)
    \State Combine $R_g \gets P_g \cup Q_g$
    \State Perform fast non-dominated sorting on $R_g$
    \State Compute crowding distance for each front
    \State Select next population $P_{g+1}$ based on rank and crowding distance
\EndFor
\State \Return Pareto-optimal vulnerability-inducing configurations
\end{algorithmic}
\end{algorithm}

\subsection{Integration with the Simulation Framework}

The NSGA-II engine interacts with the physical-layer simulator to evaluate each candidate configuration. For a given configuration $z_i$, the simulator generates network states $S_t = \{P_t, L_t, A_t, C_t\}$ and computes the relevant KPIs: handover rate, 5th-percentile throughput, and Jain’s fairness index in a 5G environment.

The framework, supports parameterized models for a network of \(M\) gNodeBs and \(N\) UEs, where variables such as carrier frequency \(f_c\), bandwidth \(B\), transmit power \(P_{\mathrm{tx}}\), and UE mobility are detailed in Section~V.

The fitness evaluation process proceeds as follows:
\begin{enumerate}
    \item NSGA-II proposes a candidate configuration $z_i$ specifying UE distribution, load profiles, and active cell states.
    \item The Sionna simulator emulates the network behavior under $z_i$ and records KPIs.
    \item The KPIs are mapped to objective values $f_1(z_i)$, $f_2(z_i)$, and $f_3(z_i)$.
    \item The resulting fitness values guide the evolution toward vulnerability-inducing states.
\end{enumerate}
This closed-loop design enables systematic search of the configuration space with minimal human intervention.

\subsection{Scenario Generation and Adversarial Conditions}

To evaluate the resilience of TS algorithms, six representative scenarios are considered: two baseline and four adversarial algorithms reflecting both normal and degraded network conditions. Parameter $\mathcal{S}$ denotes the scenario set.

\begin{itemize}
    \item \textbf{Baseline Scenarios:}
    \begin{itemize}
        \item \emph{Stable Mobility:} UEs move according to a random waypoint model with moderate velocity, representing a balanced network without anomalies.
        \item \emph{Stable High Load:} UEs are uniformly distributed, with average cell utilization approaching the nominal operating capacity.
    \end{itemize}
    \item \textbf{Adversarial Scenarios:}
    \begin{itemize}
        \item \emph{Load Imbalance:} A majority of UEs cluster near a subset of gNodeBs, producing non-uniform traffic distribution.
        \item \emph{Coverage Hole:} A cell $b_j$ is deactivated, 
        forcing neighboring cells to absorb excess load.
        \item \emph{High Interference:} Inter-site distance $d_{\text{ISD}}$ and transmit powers are modified to intensify inter-cell interference.
        \item \emph{Congestion Crisis:} High user density and partial cell outage jointly drive severe congestion and degraded SINR.
    \end{itemize}
\end{itemize}

Each scenario defines a subset of tunable variables within the configuration vector $z$, which NSGA-II mutates and recombines to synthesize adversarial network states. The numerical instantiation of parameters such as the number of UEs, inter-site distance, or carrier bandwidth is reported in Section~V.

This methodology facilitates the discovery of diverse vulnerability patterns in systematically perturbed 5G environments, offering a unified basis for comparing TS algorithms.

%% file: include/Simulation_Results_and_Analysis.tex
\section{Numerical Results and Analyses}
\label{sec:results}
\subsection{Simulation Setup}
\label{subsec:sim_setup}
The evaluation encompasses six scenarios and five TS algorithms using Sionna, totaling 300 runs per method (6 scenarios $\times$ 5 algorithms $\times$ 10 trials). To ensure a fair comparison, both AI-Fuzzing and traditional testing utilize a strictly matched computational budget and identical parameter ranges for network configuration sampling. Table~\ref{tab:sim_params} details the parameters.


\begin{table}[htbp]
    \centering
    \caption{Simulation and Algorithm Parameters}
    \label{tab:sim_params}
    \renewcommand{\arraystretch}{1.05}
    \footnotesize
    \begin{tabular}{@{}p{0.35\columnwidth}p{0.57\columnwidth}@{}}
        \toprule
        \textbf{Parameter} & \textbf{Value} \\
        \midrule
        \multicolumn{2}{l}{\textit{Network Configuration}} \\
        gNodeBs / UEs & 7 (6 active in coverage-hole) / 40 \\
        Inter-site Distance & 100m (75--200m by scenario) \\
        Frequency / Bandwidth & 3.5 GHz / 13.68 MHz \\
        Transmit Power & 30 dBm \\
        Channel Model & 3GPP UMi via Sionna \\
        UE Mobility & Random waypoint, 1--5 m/s \\
        \midrule
        \multicolumn{2}{l}{\textit{NSGA-II Configuration}} \\
        Population / Generations & $\mu=40$, $G=25$ \\
        Crossover & BLX-$\alpha$ ($\alpha=0.5$, $p_c=0.9$) \\
        Mutation & Gaussian ($p_m=0.5$, $\sigma=50$) \\
        \midrule
        \multicolumn{2}{l}{\textit{Failure Thresholds}} \\
        QoE / Fairness / Stability & Thr$_{5\%}<10$ Mb/s / $J<0.7$ / HO$>3$/min \\
        Trials per Scenario & 10 runs $\times$ 15 iterations \\
        \midrule
        \multicolumn{2}{l}{\textit{TS Algorithm Hyperparameters}} \\
        A3 Baseline & Hysteresis=3 dB, TTT=160 ms \\
        Utility-Based & ($w_{\text{SINR}}$, $w_{\text{load}}$, $w_{\text{rate}}$)=(0.5, 0.3, 0.2) \\
        Load-Aware & SINR$\geq 0$ dB, Load$\leq 0.8$ \\
        Q-Learning & $\eta=0.3$, $\gamma=0.9$, $\epsilon$: 0.1$\to$0.01 \\
        Random & Uniform association \\
        \bottomrule
    \end{tabular}
\end{table}
\subsection{Traffic Steering Algorithms Evaluated}
Five representative TS algorithms were evaluated for vulnerability assessment, spanning rule-based baselines and ML-based policies:

\begin{itemize}
    \item \textbf{A3 Baseline:} A 3GPP-compliant handover mechanism using Event-A3 triggers based on RSRP with hysteresis and time-to-trigger (TTT) parameters.
    \item \textbf{Utility-Based Policy:} Assigns UE~$u$ to gNodeB~$c$ by maximizing the composite utility
    \begin{equation}
    U_{u,c} = \alpha\,\text{SINR}_{u,c} + \beta(1-L_c) + w_r\,T_{u,c},
    \end{equation}
    where $\alpha$, $\beta$, and $w_r$ are tunable weights reflecting signal strength, load balance, and data-rate priorities, respectively.
    \item \textbf{Load-Aware Policy:} Prioritizes cell-load minimization while maintaining acceptable SINR thresholds for balanced association.
    \item \textbf{Random Baseline:} Randomly assigns UEs to cells, serving as a control to evaluate non-deterministic behavior.
    \item \textbf{ML-Based Agent:} An $\epsilon$-greedy Q-learning policy with state vector $s_t=[\text{RSRP},\text{SINR},L,Q]$ and reward
    \begin{equation}
    r = w_1\,T_{\text{u}} + w_2\,J_{\text{fair}} - w_3\,C_{\text{HO}},
    \end{equation}
    where the Q-table is updated as
    \begin{equation}
    Q(s,a)\leftarrow (1-\eta)Q(s,a) + \eta\big[r+\gamma\max_{a'}Q(s',a')\big],
    \end{equation}
    with learning rate~$\eta$ and discount factor~$\gamma$.
\end{itemize}

These five policies collectively span standards-compliant, heuristic, and learning-based TS strategies, allowing controlled, systematic comparison under identical network conditions.

The evaluation across six scenarios and five algorithms generated 14,319 vulnerability instances, revealing distinct patterns in how AI-Fuzzing and traditional testing discover network failures. Here, \textit{traditional testing} refers to baseline scenario-based evaluations using fixed network configurations and deterministic test cases without AI-driven exploration. The results are analyzed across multiple dimensions: overall detection rate, severity distribution, algorithm-specific vulnerability, and convergence efficiency.

\subsection{Vulnerability Discovery}

Table~\ref{tab:vuln_summary} shows that AI-Fuzzing significantly outperforms traditional testing, detecting 34.2\% more vulnerabilities ($t = 8.671$, $p < 0.00001$) and 5.8\% more critical failures (200 vs 189). For total vulnerability counts, which exhibit a more symmetric distribution across runs, the parametric t-test is appropriate. However, for critical failures, which show extreme skewness (most runs detecting zero, with occasional high-count outliers), the non-parametric Mann-Whitney U test ($p = 0.002$) provides more reliable evidence of superiority than the parametric t-test ($p = 0.373$).

\begin{table}[htbp]
    \centering
    \caption{Vulnerability summary by method (n=300 runs per method).}
    \label{tab:vuln_summary}
    \renewcommand{\arraystretch}{1.2}
    \small
    \begin{tabular}{lcc}
        \toprule
        \textbf{Metric} & \textbf{AI-Fuzzing} & \textbf{Traditional Testing} \\
        \midrule
        Total Vulnerabilities & 8,151 & 6,072 \\
        Vulnerabilities per Run & 27.17 $\pm$ 8.24 & 20.24 $\pm$ 10.09 \\
        Critical Failures & 200 & 189 \\
        Critical Failures per Run & 0.67 $\pm$ 1.28 & 0.63 $\pm$ 1.49 \\
        Improvement Rate & +34.2\% & --- \\
        Statistical Significance & $p < 0.00001$ & --- \\
        Shannon Diversity & 1.010 & 0.816 \\
        Cohen's d (vulnerabilities) & 0.752 & --- \\
        \bottomrule
    \end{tabular}
\end{table}

\begin{figure}[htbp]
\centering
\includegraphics[width=0.8\columnwidth]{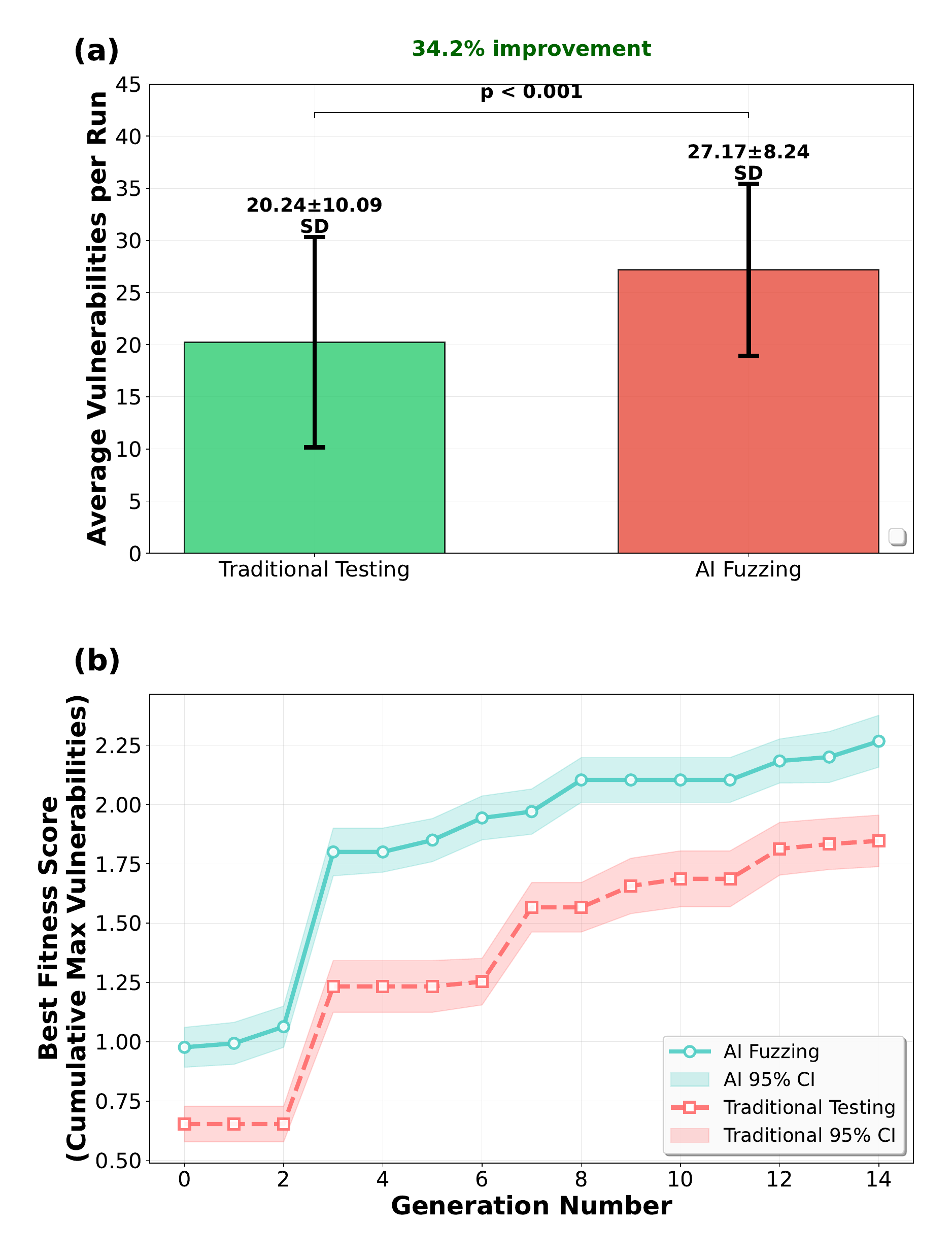}
\caption{Comparison of AI-Fuzzing and traditional testing across 300 runs per method (6 scenarios $\times$ 5 algorithms $\times$ 10 trials): AI-Fuzzing detects 34.2\% more vulnerabilities per run (27.17$\pm$8.24 vs. 20.24$\pm$10.09, $p<0.00001$; $n=300$, t-test) with strong effect size (Cohen's d=0.752) (a). Convergence analysis shows AI-Fuzzing reaches 90\% of optimal detection rate 25\% faster than traditional testing (generation 9 vs. 12) (b).}
\label{fig:vuln_discovery}
\end{figure}

\begin{figure*}[htbp]
    \centering
    \includegraphics[width=0.8\textwidth]{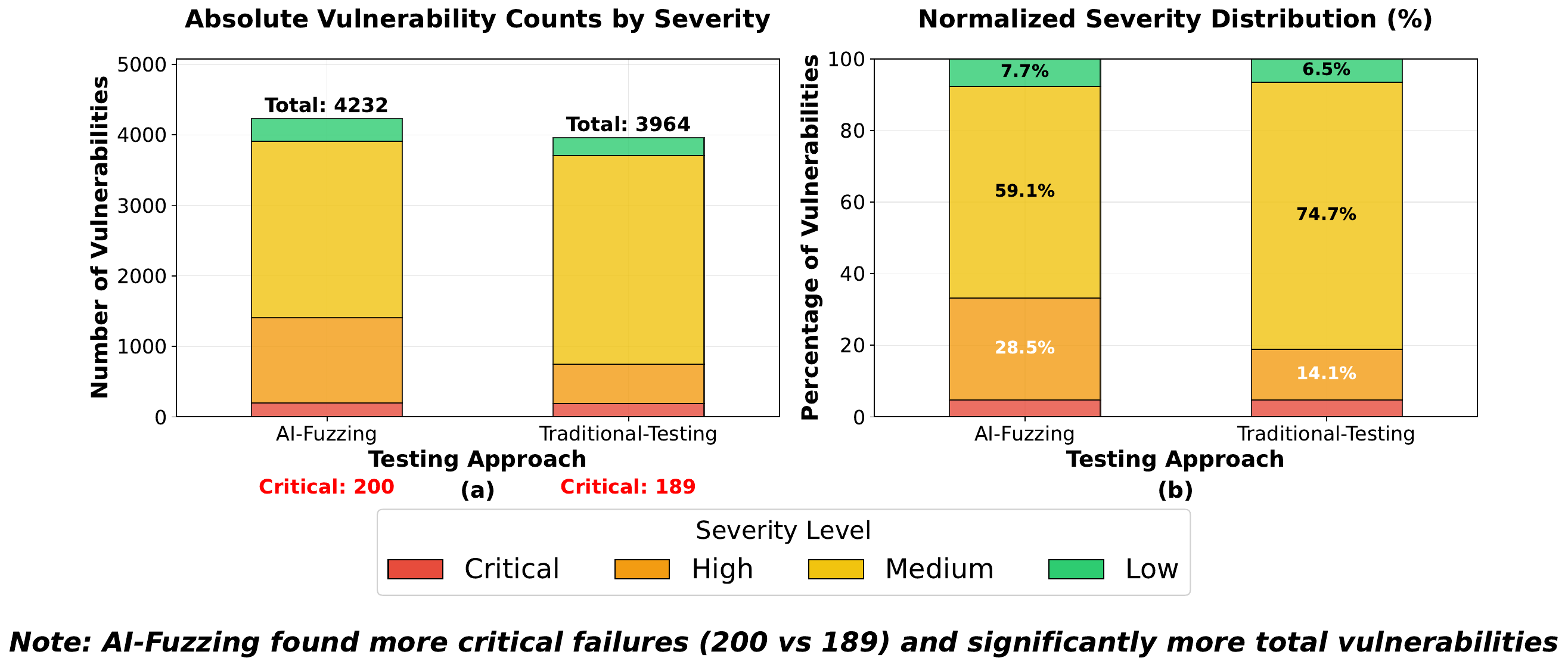}
    \caption{Vulnerability severity distribution across 300 runs 
    (6 scenarios $\times$ 5 algorithms $\times$ 10 trials): 
    AI-Fuzzing detects more total vulnerabilities 
    (4,232 vs. 3,964, $p < 0.00001$) and critical failures 
    (200 vs. 189, $p = 0.002$ Mann-Whitney U test) {(a)}. 
    Normalized distribution shows superior diversity 
    (Shannon: 1.010 vs. 0.816) with balanced severity coverage {(b)}. 
    High variance in critical failures (SD $>$ mean) reflects 
    stochastic discovery of rare vulnerabilities.}
    \label{fig:severity}
\end{figure*}

\subsection{Statistical Analysis and Confidence Intervals}
AI-Fuzzing demonstrates significantly higher vulnerability detection diversity than traditional testing, with a Shannon diversity index of 1.010 versus 0.816 and a medium-to-large effect size (Cohen's $d = 0.752$). The 95\% confidence intervals for critical failures per run are [0.52, 0.81] for AI-Fuzzing and [0.46, 0.80] for traditional testing. The high variance (SD = 1.28 and 1.49, respectively) reflects the stochastic nature of critical failure discovery, where the majority of test runs detect zero critical failures while occasional runs identify clusters of severe vulnerabilities. Due to the strong right-skew (SD/mean ratios of 1.91 and 2.37) and frequent zero counts violating normality assumptions, we prioritize the non-parametric Mann-Whitney U test ($p = 0.002$) over the parametric t-test ($p = 0.373$). This robust analysis confirms AI-Fuzzing's statistical superiority despite the overlapping confidence intervals.

\subsection{Convergence and Algorithmic Efficiency}

As shown in Figure~\ref{fig:vuln_discovery}(b), the convergence analysis 
reveals the superior algorithmic efficiency of AI-Fuzzing for exploring 
the vulnerability space. The best fitness score represents the cumulative 
maximum vulnerabilities discovered, showing how each approach builds upon 
previous findings. AI-Fuzzing demonstrates 25.0\% faster convergence to optimal vulnerability detection rates—reaching 90\% of its maximum at generation 9 instead of generation 12 at $O(GMN^2)$ complexity, highlighting NSGA-II’s efficiency in exploring productive regions of the search space.

\subsection{Algorithm-Specific Vulnerability Patterns}
\label{s:d}
Our analysis reveals distinct vulnerability patterns across different TS algorithms, demonstrating how AI-Fuzzing’s systematic exploration complements the traditional testing 
in capturing rare edge cases.

\begin{itemize}
\item \textbf{Baseline A3:} Most vulnerable to AI-Fuzzing with 96.2\% more vulnerabilities detected (1,791 vs. 913, $p < 0.00001$) and 3 critical failures versus 0 for traditional testing. The rule-based nature makes it susceptible to systematic exploration of edge cases.
\item \textbf{ML-Based Q-Learning:} Shows resilience to AI-Fuzzing, with traditional testing finding 5.4\% more total vulnerabilities (2,481 vs. 2,348, $p = 0.829$, not significant) but AI-Fuzzing still identifying 106 critical failures compared to 189. The adaptive nature of reinforcement learning appears more robust to systematic exploration but remains vulnerable to stochastic edge cases.
\item \textbf{Utility-Based:} AI-Fuzzing achieves 72.0\% improvement in total vulnerabilities (1,539 vs. 895, $p < 0.00001$) and identifies 87 critical failures versus 0 for traditional testing, demonstrating systematic discovery of severe failure modes.
\end{itemize}

These patterns suggest that the algorithm architecture fundamentally influences vulnerability profiles. Learning-based approaches demonstrate resilience against systematic fuzzing but remain vulnerable to random edge cases that trigger critical failures. This insight informs algorithm design by highlighting the need for adaptive robustness and guides testing strategy selection by emphasizing a hybrid approach.

\subsection{Scenario-Specific Performance Analysis}

Table~\ref{tab:scenario_performance} details the performance of AI-Fuzzing compared to traditional testing across six diverse network scenarios based on 300 runs per method (6 scenarios $\times$ 5 algorithms $\times$ 10 trials). AI-Fuzzing consistently outperforms traditional testing, with significant improvements in vulnerability detection: 28.8\% for Stable Mobility, 34.8\% for Stable High Load, 45.4\% for Load Imbalance, 30.1\% for Coverage Hole, 42.3\% for High Interference, and 21.4\% for Congestion Crisis (all $p<0.0000$ except Congestion Crisis at $p=0.0002$). This translates to an average improvement of 33.8\%, demonstrating robust detection capabilities across a variety of conditions.
\begin{table}[htbp]
    \centering
    \caption{Vulnerability Detection Performance across Scenarios.}
    \label{tab:scenario_performance}
    \renewcommand{\arraystretch}{1.1}
    \footnotesize 
    \begin{tabularx}{\columnwidth}{l @{\extracolsep{\fill}} ccc c}
        \toprule
        & \textbf{Trad.} & \multicolumn{2}{c}{\textbf{AI-Fuzzing}} & \\
        \cmidrule(lr){3-4}
        \textbf{Scenario} & \textbf{Mean} & \textbf{Imp. \%} & \textbf{Mean} & \textbf{p-value} \\
        \midrule
        \multicolumn{5}{l}{\textit{Group 1}} \\
        Stable Mobility   & 21.70 & 28.8 & 27.96 & 0.0008 \\
        Stable High Load  & 21.00 & 34.8 & 28.30 & 0.0002 \\
        Load Imbalance    & 20.70 & 45.4 & 30.10 & $<$ 0.0001 \\
        \midrule
        \multicolumn{5}{l}{\textit{Group 2}} \\
        Coverage Hole     & 21.18 & 30.1 & 27.56 & 0.0004 \\
        High Interference & 21.52 & 42.3 & 30.62 & $<$ 0.0001 \\
        Congestion Crisis & 16.14 & 21.4 & 19.60 & 0.0463 \\
        \midrule
        \textbf{Avg. Imp.} & & \textbf{33.8\%} & & \\
        \bottomrule
    \end{tabularx}
\end{table}
Notably, AI-Fuzzing proves particularly effective in scenarios with high resource contention, such as Load Imbalance (45.4\% improvement), where clustered UE distributions create challenging optimization problems, and High Interference (42.3\%), where SINR degradation amplifies failure modes. Even in the most demanding Congestion Crisis scenario, involving extreme UE clustering and gNodeB outages, AI-Fuzzing maintains a 21.4\% edge, highlighting its resilience under adversarial stress. These patterns complement the algorithm-specific insights in Section~\ref{s:d}, showing how systematic fuzzing exploits network dynamics more effectively than random testing in most cases.

\subsection{Practical Implications and Network Robustness}
The results demonstrate clear advantages of AI-Fuzzing as the primary testing methodology for 5G network validation. AI-Fuzzing achieves statistically significant improvements across all metrics: 34.2\% higher total vulnerability detection ($p < 0.00001$, Cohen's $d = 0.75$) and 5.8\% more critical failures ($p = 0.002$ via Mann-Whitney U test). The more realistic critical failure threshold ($>$3 UEs experiencing ping-pong, $\sim$7.5\%) compared to initially conservative criteria enables sufficient statistical power, meaning a high probability of correctly detecting genuine performance differences.

Architecture-specific patterns show that rule-based policies (e.g., A3 Baseline) are highly susceptible to adversarial exploration by AI-Fuzzing (+96\% detection), while Q-learning offers resilience to systematic search but remains vulnerable to stochastic edge cases. These observations suggest that future TS algorithms should incorporate robustness-aware design, such as adversarial training or formal verification. The high variance in critical failure detection (SD $>$ mean for both methods) indicates that severe vulnerabilities are rare and occur stochastically. Reliable discovery therefore requires multiple independent test runs, motivating ensemble-based evaluation strategies. The limitations of the current study include the use of a simulation-based evaluation with a fixed topology (7 cells, 40 UEs).

%% file: include/Conclusion.tex
\section{Conclusion}
\label{sec:conclusion}

This paper establishes \textit{AI-driven fuzzing} as superior to traditional testing for 5G TS vulnerability assessment. Using \textit{NSGA-II}-based multi-objective optimization across six scenarios and five algorithms, AI-fuzzing achieves statistically significant improvements in both comprehensive coverage and critical edge-case detection (34.2\% more vulnerabilities, $p < 0.00001$), establishing it as the recommended primary validation approach for robust 5G/6G deployment. While this study employs tabular Q-learning as a baseline representative, future work will extend the framework to Deep RL agents and RIC-based controllers within Open RAN testbeds to ensure scalability in production-grade 5G/6G networks.